\documentclass[pre,longbibliography,twocolumn,superscriptaddress]{revtex4-2}
\usepackage[utf8]{inputenc}
\usepackage{graphicx}
\usepackage{amsmath}
\usepackage{cancel}
\usepackage{color}
\usepackage{ulem}
\usepackage{setspace}
\usepackage{mathtools}
\usepackage{amssymb,latexsym}
\usepackage{lipsum}
\usepackage[svgnames]{xcolor} 
\usepackage{xcolor}

\usepackage{soul}
\usepackage[export]{adjustbox}

\definecolor{albicocca}{rgb}{0.98, 0.7, 0.2}
\definecolor{internationalorange}{rgb}{1.0, 0.31, 0.0}
\definecolor{giocolor}{RGB}{0, 100, 150}

\newcommand{\at}[2][]{#1|_{#2}}

\newcommand{\av}[1]{ \langle #1 \rangle}
\newcommand{\num}[1]{\left[ #1\right]}
\newcommand{\dens}[1]{\langle #1\rangle}
\newcommand{\parr}[1]{\left( #1\right)}

\begin{document}

\title{A pair-based approximation for simplicial contagion}  
\author{Federico Malizia}
\affiliation{Department of Physics and Astronomy,  University of Catania, 95125 Catania, Italy}
\affiliation{Network Science Institute, Northeastern University London, London E1W 1LP, United Kingdom}
\author{Luca Gallo}
\affiliation{ANETI Lab, Corvinus Institute for Advanced Studies (CIAS), Corvinus University of Budapest, 1093 Budapest, Hungary}
\author{Mattia Frasca}
\affiliation{Department of Electrical, Electronics and Computer Science Engineering, University of Catania, 95125 Catania, Italy}
\author{Istv\'an Z. Kiss}
\affiliation{Network Science Institute, Northeastern University London, London E1W 1LP, United Kingdom}
\author{Vito Latora}
\affiliation{Department of Physics and Astronomy,  University of Catania, 95125 Catania, Italy}
\affiliation{INFN Sezione di Catania, Via S. Sofia, 64, 95125 Catania, Italy}
\affiliation{School of Mathematical Sciences, Queen Mary University of London, London E1 4NS, UK}
\affiliation{Complexity Science Hub Vienna, A-1080 Vienna, Austria}
\author{Giovanni Russo}
\affiliation{Department of Mathematics and Computer Science, University of Catania, 95125 Catania, Italy}
\date{\today}

\begin{abstract}

Higher-order interactions play an important role in complex contagion processes. Mean-field approximations have been used to characterize the onset of spreading in the presence of group interactions. 
However, individual-based mean-field models are unable to capture correlations between different subsets of nodes, which can significantly influence the dynamics of a contagion process.
In this paper, we introduce a pair-based mean-field approximation that allows to study the dynamics of a SIS model on simplicial complexes 
by taking into account correlations at the level of pairs of nodes.
Compared to individual-based mean-field approaches, the proposed approximation yields more accurate predictions of the dynamics of contagion processes on simplicial complexes. 
Specifically, the pair-based mean-field approximation provides higher accuracy in predicting the extent of the region of bistability, the type of transition from  disease-free to endemic state, and the average time evolution of the fraction of infected individuals. Crucially, the pair-based approximation correctly predicts that the onset of the epidemic outbreak in simplicial complexes depends on the strength of higher-order interactions.
Overall, our findings highlight the importance of accounting for pair correlations when investigating contagion processes in the presence of higher-order interactions.


\end{abstract}

\maketitle

\section{Introduction}
Complex networks have been used to describe a large variety of dynamical processes involving interacting units, such as epidemic spreading \cite{kiss2017mathematics, pastor2015epidemic}, social contagion \cite{gomez2016explosive}, random walks \cite{noh2004random}, and synchronization \cite{gomez2008entropy}, among many others.
However, networks have inherent limitations as they can only capture interactions between pairs of units \cite{boccaletti2006complex,latora2017complex}. Consequently, they cannot be employed to study complex systems characterized by interactions occurring in groups of three or more units \cite{battiston2020networks}. In such cases, more sophisticated mathematical structures, such as hypergraphs \cite{berge1984hypergraphs} and simplicial complexes \cite{atkin1972cohomology}, are required. The inclusion of higher-order interactions has proved that these interactions give rise to novel collective phenomena in various dynamical processes \cite{carletti2020dynamical}, including diffusion \cite{carletti2020random, schaub2020random}, percolation \cite{sun2021higher},  synchronization \cite{skardal2019abrupt,lucas2020multiorder, gambuzza2021stability}, and evolutionary games \cite{alvarez2021evolutionary}.

In particular, the framework of higher-order networks has proven to be essential in modeling social contagion processes, such as opinion formation, rumor spreading, or the adoption of novelties, where the exposure to multiple sources is needed to trigger the transmission \cite{centola2007complex}. Recent studies of social contagion in hypergraphs \cite{de2020social} and simplicial complexes \cite{iacopini2019simplicial} have revealed that higher-order interactions may radically change the characteristics of the spreading process. In the presence of higher-order interactions, the transition to an endemic state in SIS models can become discontinuous and a bistable regime where a disease-free and an endemic state co-exist can appear \cite{iacopini2019simplicial,de2020social}. This behavior is closely connected to the microscopic organization of higher-order interactions as diverse structures may exhibit different types of transitions \cite{malizia2023hyperedge}.

Most models for the analysis of social contagion processes with higher-order interactions rely on individual-based mean-field approaches, assuming that the individuals are homogeneously mixed and interact with each other at random. However, this assumption may lead to an oversimplified approximation, as it overlooks the dynamic correlations that arise within the underlying structure. For example, infected individuals are more likely to come into contact with other infected individuals \cite{pastor2015epidemic}. 
In the context of complex networks, to obtain a more precise description of the spreading process, one can consider more sophisticated approximations. 
For instance, pair-based models can be employed \cite{keeling1999effects,sharkey2015exact,frasca2016discrete,malizia2022individual}. 
In these models, the system dynamics is not characterized at the level of nodes but at the level of node pairs.
That is, they do not follow the temporal evolution of the expected number of individuals in a given state, they analyze instead how the expected number of edges in a given state evolves in time.
However, deriving pair-based mean-field models for spreading processes on structures with higher-order interactions remains challenging. 



In this work, we focus on the complex contagion process introduced in Ref.~\cite{iacopini2019simplicial}, namely an SIS process in the presence of three-body simplicial interactions. We develop a mean-field model of the process based on a description at the level of node pairs, and we 
show that it reproduces  
the results of stochastic simulations of the process 
better than an individual-based mean-field model~\cite{iacopini2019simplicial}.

Recently, several novel approaches have been proposed to study and capture the emerging behavior in contagion processes in the presence of higher-order interactions. These methods include quenched group-based model \cite{burgio2021network}, group-based approximate master equations \cite{st2021master}, quenched pair approximation \cite{matamalas2020abrupt} and triadic approximation \cite{burgio2023triadic}.
%
In particular, the authors of Ref.~\cite{matamalas2020abrupt} have proposed a discrete-time model relying on a Markov-chain. The system dynamics is expressed in terms of joint probabilities of the microscopic states of the links and the nodes, and is thus described by a set of $N+L$ master equations, where $N$ is the number of nodes and $L$ the number of links. 
In our work, instead, we consider a framework that is continuous-time and with a number of governing equations that does not depend on the number of nodes and links.
%
%
%
%
More recently, 
the authors of Ref.~\cite{burgio2023triadic} 
have introduced a 
mean-field triadic approximation, which allows to describe higher-order contagion processes by associating state variables not only to nodes and pairs of nodes but also to groups of three nodes.
This model correctly predicts key features of complex contagion, including the epidemic threshold and the stationary density of spreaders, and works for both simplicial complexes and hypergraphs. 
The only limitation of this approach is that it relies on the assumption that two three-body interactions (hyperedges or simplices) share at most one node.
This can be a too strong assumption in structures with high clustering.
The mean-field approximation we introduce in this paper stops at the level of pairs, but it allows one to properly account for correlations arising in
motifs involving three and four nodes within a simplicial complex.
Despite being limited to the case where higher-order interactions are represented by simplicial complexes our work provides a complementary approach to existing models, 
and can provide further insights into contagion processes in presence of higher-order interactions.

The paper is organized as follows. 
In Sec.~\ref{sec:preliminaries} we provide 
the basic ideas and methods behind individual-based and  pair-based mean-field approximations of the SIS model on networks.  
In Sec.~\ref{sec:pair-based-model} we focus instead on the case of the SIS model on simplicial complexes. We review the individual-based mean-field approximation proposed in Ref.~\cite{iacopini2019simplicial}, and we introduce our mean-field pair-based approximation 
of the simplicial SIS model. 

In Sec.~\ref{sec:results},  we show the advantages of a pair-based mean-field approximation compared to an individual-based one in reproducing the results of stochastic simulations of the simplicial SIS model. Furthermore, in the same section we derive an analytical  expression of the epidemic threshold in the pair-based approximation.
%
Finally, in Sec.~\ref{sec:conclusions} we summarize and discuss the main results obtained.

\section{Mean-field SIS models}
\label{sec:preliminaries}

In this section, we focus on the case of simple contagions, i.e. processes where transmission occurs exclusively through pairwise interactions. We discuss the two standard mean-field approximations for the SIS model on networks. 
Specifically, we describe the individual-based and the pair-based mean-field, highlighting the mathematical approach used to derive mean-field models at the individual- and pair-based levels. 
This will lay the groundwork for revisiting the individual-based approximation for the simplicial SIS model studied in Ref.~\cite{iacopini2019simplicial} and for introducing our pair-based approximation in Sec.~\ref{sec:pair-based-simplicial-model}.


\subsection{Individual-based mean-field SIS model}

We begin by characterizing the processes governing the transition of an individual from one state to another in the standard SIS model. In this model, an individual can be either susceptible (S) or infected/infectious (I), and can transit from one compartment to another as infection or recovery takes place. A susceptible individual (S) becomes infected (S$\rightarrow$I) after interacting with an infectious individual (I), who acts a mediator of the transition. The infection mechanism is, hence, a two-body nonlinear process. Infected individuals (I) recover after a given period of time, becoming once again susceptible (I$\rightarrow$S). Contrarily to the infection mechanism, recovery is described by a one-body linear process. Formally, we can express these transitions in terms of two kinetic equations   
\begin{equation}
\begin{array}{rcl}
     S + I & \overset{\beta}{\rightarrow} & I + I \\
     I & \overset{\mu}{\rightarrow} & S
\end{array},
\label{eq:standard_SIS_transitions}
\end{equation}
where $\beta$ and $\mu$ are the transition rates for the infection and the recovery processes, respectively.

\begin{figure*}[t!]
    \centering
    \includegraphics[width=\linewidth]{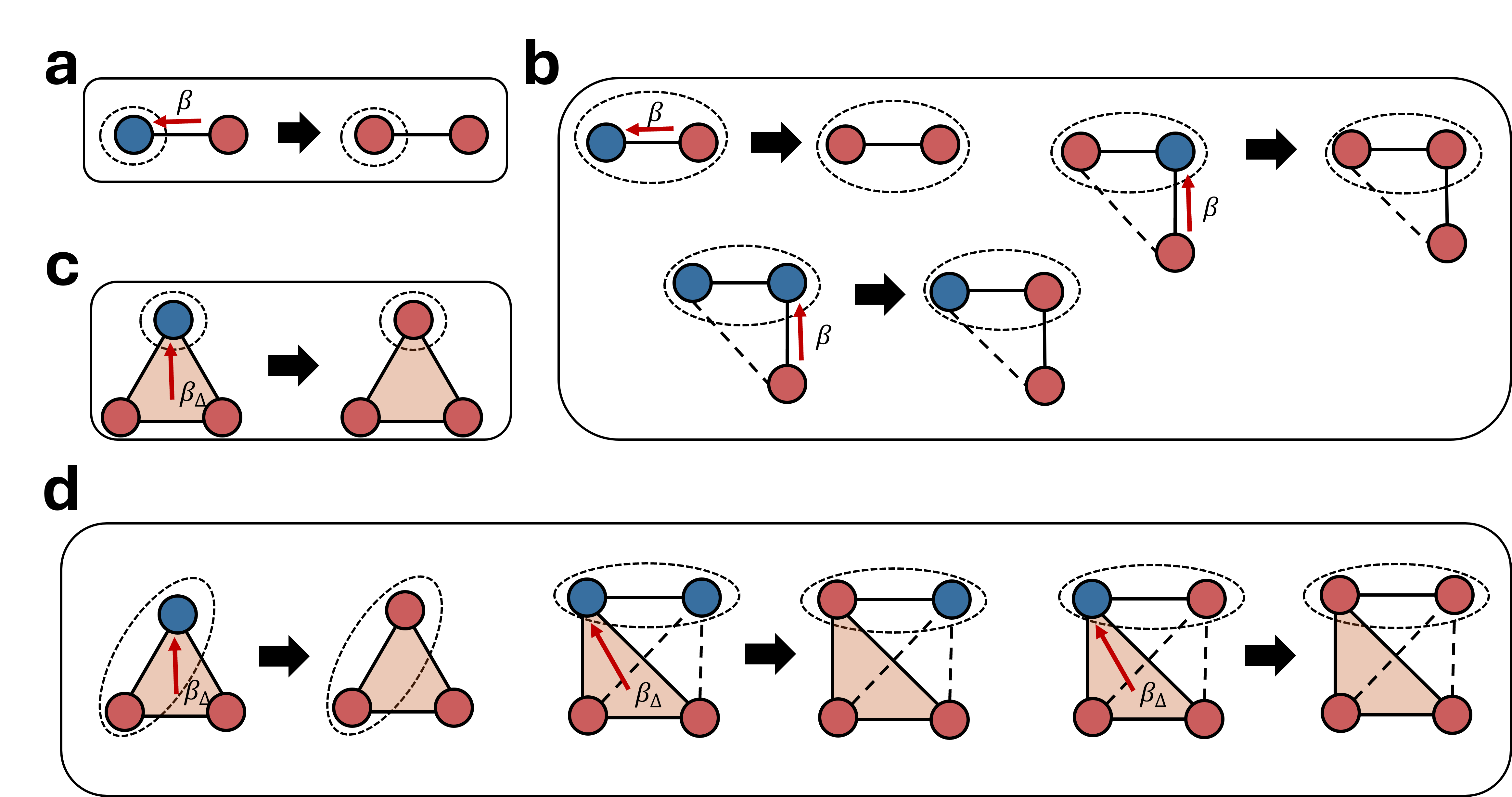}
    \caption{Pictorial representation of infection processes among susceptible (in blue) and infected (in red) individuals in SIS models. (a) Infection of a node connected to an infected node through a link. This is the only infection process occurring in the individual-based SIS model. (b) Infection of a pair of nodes. In a pair-based description of contagion in a SIS model, the disease can be transmitted to one of the nodes of a pair by the other node of the pair or by other infected nodes not belonging to the pair. (c) Simplicial infection of a node by a group interaction with two infected nodes. (d) Infection of a pair of nodes in the presence of group interactions. The infection processes depicted in panel (d) arise in the pair-based description of the simplicial SIS model.}
\label{fig:transitions_tables}
\end{figure*}

The dynamics of the SIS model on network can be investigated using different approaches \cite{pastor2015epidemic,kiss2017mathematics}. 
A commonly employed strategy is to perform stochastic simulations of the kinetic equations~\eqref{eq:standard_SIS_transitions}. 
However, this method has some limitations. 
Firstly, it is computationally expensive for large networks. 
Also, it requires a significant amount of data, as both the contact network and the states of all nodes need to be known empirically. 
Finally, this approach does not facilitate the understanding of the relationship between the model parameters and the emerging collective behavior \cite{malizia2022individual}.
For this reason, it is common to use a deterministic representation of the process that focuses on population-level quantities, such as the density or the expected number of individuals in a given state. 
This deterministic approach assumes homogeneous mixing: interactions between individuals are treated as uniform and independent of the network structure, so that each individual can be considered statistically equivalent to any other. 
This hypothesis allows us to describe the dynamics of the system in terms of a so-called mean-field model. 

The SIS process in homogeneous mixing is described by the following equations
\begin{equation}
\begin{array}{ccc}
     \dot{\langle S\rangle} &=& \mu\langle I\rangle-\beta k\langle SI\rangle \\
     \dot{\langle I\rangle} &=& -\mu\langle I\rangle +\beta k\langle SI\rangle 
\end{array},
\label{eq:standard_SIS_MF}
\end{equation}
where $\langle S\rangle$ and $\langle I\rangle$ represent the densities of susceptible and infected individuals, respectively, $\langle SI\rangle$ is the density of edges in the state $(S,I)$, i.e., a susceptible individual in contact with an infectious one (note that pairs are unordered, meaning that we do not distinguish $(S,I)$ from $(I,S)$), while $k$ is the average number of contacts.
The densities of individuals in state S and I are given by $\langle S\rangle = \num{S}/N$ and $\langle I\rangle = \num{I}/N$, where $\num{S}$ and $\num{I}$ are the expected number of susceptible and infected individuals, respectively, and $N=[S]+[I]$ is the total number of individuals in the population. 
Consequently, $\langle S\rangle$ and $\langle I\rangle$ are not independent and 
satisfy the conservation relation $\langle S\rangle+\langle I\rangle=1$. As can be easily checked, the terms on the right hand sides of Eqs.~\eqref{eq:standard_SIS_MF}
sum to zero, so that $\dot{\av{S}} + \dot{\av{I}} = 0$.
The density $\langle SI\rangle$ is defined as $\langle SI\rangle = \num{SI}/kN$, where $\num{SI}$ represents the expected number of edges in state $(S,I)$. 
Note that $kN=2L$, where $L$ is the total number of edges within the network, assuming it to be undirected. 
The two terms in the equation for $\langle I\rangle$ represent the decrease in the density of infectious individuals due to recoveries, and the increase due to infections, respectively. 
The first depends on the transition rate $\mu$ and on the density of infected individuals, while the second is given by the transition rate $\beta$, the average degree $k$ and the edge density $\langle SI\rangle$. 
The transition from susceptible to infected through the interaction with an infected individual is depicted in panel (a) of Fig.~\ref{fig:transitions_tables}, from which one can derive the interaction terms appearing in \eqref{eq:standard_SIS_MF}.

System~\eqref{eq:standard_SIS_MF} is exact but not closed, as the equations governing the dynamics of $\langle S\rangle$ and $\langle I\rangle$ depend on the quantity $\langle SI\rangle$. 
To obtain a closed system of equations, one may apply the law of mass action, assuming statistical independence at the level of individuals. 
This means to assume that infected individuals are randomly distributed in the network, so the probability that a neighbor is infected is simply given by $\langle I\rangle$ and does not depend on the state of the node itself. 
Under this assumption, we can approximate the fraction of edges in state $(S,I)$ as 
\begin{equation}\label{eq:si_closure}
    \langle SI\rangle \approx \langle S\rangle\langle I\rangle,
\end{equation}
and substituting this expression in Eqs.~\eqref{eq:standard_SIS_MF}, we obtain a closed form for the mean-field SIS model
\begin{equation}
\begin{array}{ccc}
     \dot{\langle S\rangle} &=& \mu\langle I\rangle -\beta k\langle S\rangle\langle I\rangle\\
     \dot{\langle I\rangle} &=& -\mu\langle I\rangle +\beta k\langle S\rangle\langle I\rangle \\
\end{array}.
\label{eq:standard_SIS_MF_densities}
\end{equation}

\subsection{Pair-based mean-field SIS model}
\label{sec:pair-based-standard}
Assuming that infected individuals are randomly distributed in the network may be a too rough approximation, as it does not account for the dynamic correlations that exist within the contact network (e.g., infected nodes are more likely to come into contact with other infected nodes)~\cite{kiss2017mathematics}. 
To provide a more accurate description of the SIS dynamics on a network, we can consider a pair-based model, incorporating these dynamic correlations. 
In practice, one has to characterize the system dynamics at the level of pairs of nodes, describing how the expected number of edges in a given state evolves in time. 
With reference to Fig.~\ref{fig:transitions_tables},
the equations describing the SIS process at the level of pairs of nodes are
\begin{equation}
\label{eq:pairwise_SIS_MF_exact}
\begin{array}{ll}
\dot{\dens{S}} =& \mu \dens{I} - \beta k\dens{SI}\\
\dot{\dens{I}} =& - \mu \dens{I} + \beta k \dens{SI} \\
\dot{\dens{SS}} =& 2\mu\dens{SI} - 2\beta(k-1)\dens{SSI}\\
\dot{\dens{SI}} =& \mu \dens{II} - \mu \dens{SI} + \beta(k-1)\dens{SSI} \\& - \beta(k-1)\dens{ISI} - \beta \dens{SI} \\
\dot{\dens{II}} =& -2\mu\dens{II} + 2\beta(k-1)\dens{ISI} + 2\beta\dens{SI}
\end{array},
\end{equation}
where $\dens{ISI}$ represents the density of (both open and closed, and unordered) triplets in state $(I,S,I)$ , and is defined as 
$\dens{ISI}=\num{ISI}/(k(k-1)N)$, with $\num{ISI}$ being the expected number of triplets in that state \footnote{Note that factors 2 in the equations come from the fact that pairs of nodes are unordered \cite{malizia2022individual}}. 
Note that the term $k(k-1)$ corresponds to the average number of triplets connected to each node.
Indeed a node with $k$ neighbors is at the center of $k(k-1)/2$ unordered triples. 
$\langle S\rangle$ and $\langle I\rangle$ can be obtained from $\langle SS\rangle$, $\langle SI\rangle$ and $\langle II\rangle$ through marginalization, namely $\langle S\rangle=\langle SS\rangle+\langle SI\rangle$ and $\langle I\rangle=\langle SI\rangle+\langle II\rangle$. 
Additionally, $\langle SS\rangle$, $\langle SI\rangle$ and $\langle II\rangle$ are not independent, as they are linked through the conservation relation $\langle SS\rangle + 2\langle SI\rangle + \langle II\rangle =1$.
It can be easily checked that,  similarly to Eqs.~\eqref{eq:standard_SIS_MF}, also Eqs.~\eqref{eq:pairwise_SIS_MF_exact}  satisfy $\dot{\dens{S}} + \dot{\dens{I}} = 0$. Moreover, Eqs.~\eqref{eq:pairwise_SIS_MF_exact}   
satisfy $\dot{\dens{SS}} + \dot{\dens{SI}} = \dot{\dens{S}}$, $\dot{\dens{SI}}+\dot{\dens{II}} = \dot{\dens{I}}$, where we used the relation 
${\dens{SSI}}+{\dens{ISI}} = {\dens{SI}}$.

Note that, with respect to the individual-based approximation, the pair-based SIS model contains a higher number of terms.
This reflects the higher number of possible states and transitions that a pair can undergo compared to a single node.

To illustrate the terms in Eqs.~(\ref{eq:pairwise_SIS_MF_exact}), let us focus on the equation governing the dynamics of $\dens{SI}$. 
The first two terms are related to recoveries: the first represents the increase in the density $\dens{SI}$ due to the recovery of one of the nodes of the pairs in state $(I,I)$; 
the second models the decrease in $\dens{SI}$ due to the recovery of the infected nodes of the pairs in state $(S,I)$, i.e., the transition from $(S,I)$ to $(S,S)$. 
The remaining three terms encode the transitions due to infections, as graphically represented in panel (b) of Fig.~\ref{fig:transitions_tables}. 
In particular, the first term captures the increase in $\dens{SI}$ due to the infection of one of the nodes of the pairs in $(S,S)$, while the second represents the decrease in $\dens{SI}$ due to the infection of the susceptible nodes of the pairs in state $(S,I)$. 
Both transitions are due to the interactions with a third (infectious) node, among the $(k-1)$ remaining neighbors of the susceptible nodes.
Note also that the three nodes involved in the process can be arranged in two possible motifs, namely an open triangle, i.e., a wedge, or a closed one. 
This point will turn out to be crucial when considering the closure of the system. 
Finally, the last term corresponds to the decrease of $\dens{SI}$ due to the infection of the susceptible nodes in the pairs in state $(S,I)$. 
In this case, however, the transition is due to the infected nodes of the pair, and not to a third node.

System \eqref{eq:pairwise_SIS_MF_exact} is exact but not closed, as the equations governing the dynamics of $\langle SS\rangle$, $\langle SI\rangle$ and $\av{II}$ depend on the quantities $\av{ISI}$ and $\av{SSI}$. Therefore, various approximations of the densities of the triplets have been proposed to close the system at the pair level. Here, we use one of the most common and studied closures \cite{kiss2017mathematics}, which consists of writing the probability that a triplet is in a given state as the product of the probabilities that its edges are in a certain state, normalized by the probabilities that the nodes in common to the pairs of edges are in a particular state.
A key point of this closure is to distinguish between open and closed triangles. In the first case, the three nodes, namely $x$, $y$, and $z$ form a wedge, that is, node $y$ is linked to nodes $x$ and $z$, but $x$ and $z$ are not connected. In the second case, there is a link also between $x$ and $z$ (we call this configuration, shortly, a triangle).

Let us first consider the case in which three nodes $x$, $y$, and $z$ forming a wedge are in states A, B, and C, respectively.  We can write the density of wedges in state $(A,B,C)$ as 
\begin{equation}\label{eq:wedge_closure}
    \begin{array}{l}
         \av{ABC^\wedge} \approx \dfrac{\av{AB}\av{BC}}{\av{B}}, \\
    \end{array}
\end{equation}
where $\av{AB}=P(AB)$ and $\av{BC}=P(BC)$ are the probabilities that links $(x,y)$ and $(y,z)$ are in states $(A,B)$ and $(B,C)$, respectively, while $\av{B}$ is the probability that node $y$ is in state B. 
This formula can be obtained as follows.
$\av{ABC^\wedge}$ is given by
\[
    \av{ABC^\wedge} = P(AB\cap BC) = P(BC) P(AB|BC)
\]
where $P(AB|BC)$ denotes the conditional probability that $x$ and $y$ are in states $A$ and $B$, given that $y$ and $z$ are in states $B$ and $C$. 
Since $x$ and $z$ are not connected, we can approximate the conditional probability as
\[
    P(AB|BC) \approx P(AB|B) = P(AB)/P(B)
\]
Combining these two relations we obtain Eq.~\eqref{eq:wedge_closure}.
When the network has few short cycles, i.e., few closed triangles, one can use $\av{ABC}\simeq\av{ABC^\wedge}$ and Eq.~\eqref{eq:wedge_closure} that, in this case, give an accurate approximation of the probability that a triplet of nodes is in state $(A,B,C)$. 
However, this closure fails to describe the epidemic dynamics on networks where the number of closed triplets is not negligible \cite{kiss2017mathematics,keeling1999effects}, as it overlooks the correlation between nodes $x$ and $z$. 

Let us now consider the case when the three nodes form a closed triangle. Here, to approximate the density of triangles in state $(A,B,C)$ the so-called Kirkwood superposition~\cite{kirkwood1935statistical}, namely
\begin{equation}\label{eq:kirkwood_closure}
    \begin{array}{l}
     \av{ABC^\Delta} \approx \dfrac{\av{AB}\av{BC}\av{AC}}{\av{A}\av{B}\av{C}}
    \end{array}.
\end{equation} 
can be used~\footnote{The formula can be understood as follows:
$P\left( AB\cap BC\cap CA\right) =P\left(  AB\cap BC| CA\right) P\left( CA\right)$. Now
$P\left(  AB\cap BC| CA\right) =P\left(  AB| BC\cap CA\right) \cdot P\left(  BC| CA\right) $. We can write $P\left(  AB| BC\cap CA\right) =\dfrac{P\left( AB\right) }{P\left( A\right) P\left( B\right) }$, and use the approximation
$P\left(  AB| BC\cap CA\right) =\dfrac{P\left( AB\right) }{P\left( A\right) P\left( B\right) }$. Substituting in the previous expression one has 
$ P(AB\cap BC|CA) =\dfrac{P\left( AB\right) }{P\left( A\right) P\left( B\right) }\cdot \dfrac{P\left( BC\right) }{P\left( C\right) }$ from which it follows
$P\left( AB\cap BC\cap CA\right) ={P( AB) P(BC) P( CA) }/{(  P(A) P( B) P(C) )}$}. Note that, if we assume the dynamics of nodes $x$ and $z$ to be uncorrelated, i.e., $\av{AC}=\av{A}\av{C}$, the approximation for $\av{ABC^\Delta}$ recovers the one for $\av{ABC^\wedge}$.

The final step is to calculate the probability that a generic triplet of nodes is in state $(A,B,C)$, namely $\av{ABC}$, as a function of $\av{ABC^\wedge}$ and $\av{ABC^\Delta}$. 
To this aim, we consider the global clustering coefficient $\phi \in [0,1]$, representing the fraction of closed triangles over all triplets in the structure, that we can calculate in a statistical meaning as 
\begin{equation}
    \begin{array}{l}
\displaystyle \phi = \dfrac{3 \mathbb{E}(\mathcal{T})}{N k(k-1)},
    \end{array}
    \label{eq:clusteringcoeff}
\end{equation}
where $N$ is the number of nodes, $k$ is the average number of links, and $\mathbb{E}(\mathcal{T})$ represents the expected number of triangles, which depends on the structure considered. 
At this point, we can calculate $\av{ABC}$ as \cite{keeling1999effects,kiss2017mathematics}:
\begin{equation}\label{eq:closure_used}
    \av{ABC} = (1-\phi)\av{ABC^\wedge} + \phi \av{ABC^\Delta}.
\end{equation}
Note that the densities $\dens{ABC^\wedge}$ and $\dens{ABC^\Delta}$ are defined as $\dens{ABC^\wedge} = \num{ABC^\wedge}/(k(k-1)(1-\phi)N)$ and $\dens{ABC^\Delta} = \num{ABC^\Delta}/(k(k-1)\phi N)$, where $\num{ABC^\wedge}$ and $\num{ABC^\Delta}$ are the expected number of wedges and of triangles in state $(A,B,C)$, respectively. 

Summing up, we consider Eq.~\eqref{eq:closure_used} to close the compartmental models at the pair level. 
In particular, we will approximate the densities $\dens{SSI}$ and $\dens{ISI}$ as:
\begin{equation}\label{eq:closed_triples}
    \begin{array}{l}
         \av{SSI} \approx \parr{1-\phi}\dfrac{\av{SS}\av{SI}}{\av{S}} + \phi\dfrac{\av{SS}\av{SI}^2}{\av{S}^2\av{I}}\\
         \av{ISI} \approx \parr{1-\phi}\dfrac{\av{SI}^2}{\av{S}} + \phi\dfrac{\av{SI}^2\av{II}}{\av{S}\av{I}^2}
    \end{array}.
\end{equation}
Eqs.~\eqref{eq:pairwise_SIS_MF_exact} along with the expressions \eqref{eq:closed_triples} constitute a closed approximation for the pair-based mean-field SIS model.

\section{MEAN-FIELD SIMPLICIAL SIS MODELS}
\label{sec:pair-based-model}

In this section, we consider the case of complex contagion. We focus on the simplicial contagion model, and  
we begin by describing an individual-based approximation of such model \cite{iacopini2019simplicial}. 
Then, we introduce our novel pair-based mean-field approximation of simplicial contagion that extends the pair-based approximation on networks discussed in Sec.~\ref{sec:pair-based-standard} to the case of simplicial complexes.

\subsection{Individual-based mean-field simplicial SIS model}
\label{sec:individual-based}

We now discuss the individual-based mean-field simplicial SIS model presented in \cite{iacopini2019simplicial}. 
We first describe the processes ruling the transition of an individual from one state to another.
Compared to the SIS process on networks (see \eqref{eq:standard_SIS_transitions}), there is a further way in which an individual can transit from one compartment to another. 
Specifically, a susceptible individual (S) can become infected (S$\rightarrow$I) through a three-body interaction, in which the other two individuals are infectious.
In this higher-order interaction, two infected individuals at the same time act as mediators of the transition. 
We can represent the processes of the simplicial SIS model in terms of kinetic equations   
\begin{equation}
\begin{array}{rcl}
     S + I & \overset{\beta}{\rightarrow} & I + I \\
     S + I + I & \overset{\beta_\Delta}{\rightarrow} & I + I + I\\
     I & \overset{\mu}{\rightarrow} & S
\end{array},
\label{eq:simplicial_SIS_transitions}
\end{equation} 
where $\beta$ and $\beta_\Delta$ are the transition rates for the two-body and the three-body infection process, respectively, while $\mu$ is the recovery rate. 

We now derive a model that describes the system dynamics in terms of population-level quantities. 
In this framework, we consider each three-body interaction to also include all possible pairwise interactions among the three nodes. This represents the essential feature of simplicial complexes, known as their inclusion property or downward closure \cite{salnikov2018simplicial}, which makes them a very special case of hypergraphs.
Under the homogeneous mixing hypothesis, the exact equations for the simplicial SIS model are given by  
\begin{equation}
    \begin{array}{ll}
         \dot{\dens{S}} =& \mu\dens{I} - \beta k\dens{SI} - \beta_\Delta k(k-1)\phi\delta\dens{ISI^\Delta} \\
         \dot{\dens{I}} =& -\mu\dens{I} + \beta k\dens{SI} + \beta_\Delta k(k-1)\phi\delta\dens{ISI^\Delta}
    \end{array},
    \label{eq:individual_SIS_simplagion_order1}
\end{equation}
where $\delta\in [0,1]$ is the fraction of triangles that are effectively $2$-simplices, i.e., they represent a three-body interaction. 
Compared to Eqs.~\eqref{eq:standard_SIS_MF}, these equations present an additional term, i.e., $\beta_\Delta k(k-1)\phi\delta\dens{ISI^\Delta}/2$, which takes into account the infections of susceptible nodes due to the simultaneous interaction with two infected individuals, as shown in panel (c) of Fig.~\ref{fig:transitions_tables}.

System \eqref{eq:individual_SIS_simplagion_order1} may be closed by applying the law of mass action, starting from the assumption that infected individuals are randomly distributed in the simplicial complex \cite{iacopini2019simplicial}. 
Accordingly, the density $\dens{SI}$ is approximated as in Eq.~\eqref{eq:si_closure}, while $\dens{ISI^\Delta}$ is approximated as
\begin{equation}
    \dens{ISI^\Delta} \approx \dens{I}\dens{S}\dens{I}.
\end{equation}

We can therefore write the individual-based mean-field model for the simplicial SIS model as
%
\begin{equation}
    \begin{array}{ll}
         \dot{\dens{S}} =& \mu\dens{I} - \beta k\dens{S}\dens{I} - \beta_\Delta k_\Delta\dens{S}\dens{I}^2 \\
         \dot{\dens{I}} =& -\mu\dens{I} + \beta k\dens{S}\dens{I} + \beta_\Delta k_\Delta\dens{S}\dens{I}^2
    \end{array},
    \label{eq:individual_SIS_simplagion_order1_closed}
\end{equation}
where $k_\Delta = k(k-1)\phi\delta/2$ represents the average number of $2$-simplices connected to each node. 

The system can be analytically investigated in terms of two parameters, $\lambda = k\beta/\mu$ and $\lambda_\Delta = k_\Delta\beta_\Delta /\mu$, representing the rescaled infectivity on 1-simplices, i.e., links, and on 2-simplices, resepctively. 
In particular, the steady-state solutions of Eqs.~\eqref{eq:individual_SIS_simplagion_order1_closed} and their stability can be studied as a function of $\lambda$ and $\lambda_\Delta$. 
The analysis \cite{iacopini2019simplicial} shows that, when the higher-order interactions are weak, i.e., when $\lambda_\Delta < 1$, the system behaves similarly to the SIS model on networks, described by Eq.~\eqref{eq:standard_SIS_MF}: for $\lambda < 1$ the disease-free equilibrium $\dens{I}^* = 0$ is the only solution, while for $\lambda > 1$ a stable endemic state $\dens{I}^* \neq 0$ exists.

In addition, the phase transition at $\lambda = 1$ is continuous. 
Instead, when the higher-order interactions are strong, i.e., $\lambda_\Delta > 1$, the system shows a different behavior. 
For $\lambda$ smaller than $\lambda_c = 2\sqrt{\lambda_\Delta} - \lambda_\Delta$, only the disease-free equilibrium exists; for $\lambda_c < \lambda < 1$, the disease-free equilibrium and an endemic state coexist in a bistable regime, with the initial fraction of infected individuals determining whether the system reaches one equilibrium or the other: If the initial fraction of infected individuals is larger than a critical mass \cite{iacopini2019simplicial}, the system evolves towards the endemic state; if not, the spreading cannot be sustained, and the system evolves towards the disease-free equilibrium. 
Finally, for $\lambda > 1$, the endemic state is the only stable equilibrium. 
Differently from the SIS model on networks, when $\lambda_\Delta > 1$ the phase transitions occurring at $\lambda = \lambda_c$ and $\lambda = 1$ are discontinuous.

\subsection{Pair-based mean-field simplicial SIS model}\label{sec:pair-based-simplicial-model}

\begin{figure}[t!]
    \centering
    \includegraphics[width=\linewidth]{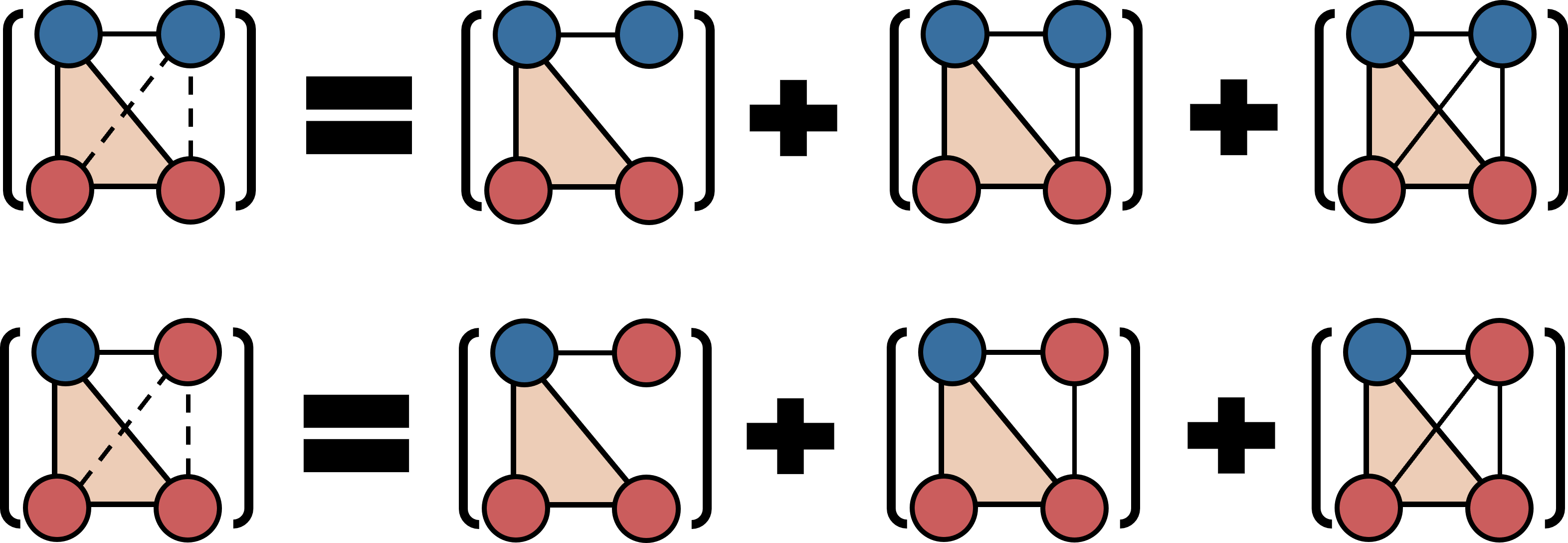}
    \caption{Graphical representation of the three possible microscopical configurations of four-node motif states $(I,I,S,S)$ (on top) and $(I,I,S,I)$ (bottom) that contain a single 2-simplex at the state (I,S,I), as defined in Eq.~\eqref{eq:4node_motif_relation}. Square brackets refer to the expected number of the  configurations containing only one 2-simplex, which are obtained through the relation $\num{ABC^\Delta D} = Nk_\Delta (k-2) [(1-\phi)^2 \av{ABC^\Delta D}_0 + \phi(1-\phi) \av{ABC^\Delta D}_1 + \phi^2\av{ABC^\Delta D}_2]$.}
    
\label{fig:4motifs_state}
\end{figure}

In this section, we derive the pair-based mean-field simplicial SIS model. 
To this aim, we have to account for all ways a pair of nodes in a simplicial complex can transit from one state to another. 
Besides those related to recoveries and two-body infections (see Sec.\ref{sec:preliminaries}), the transitions determined by three-body infections (illustrated in panel (d) of Fig.~\ref{fig:transitions_tables}) should also be considered. 
Remarkably, two of these transitions depend on the state of 4-node motifs, and occur when the $2$-simplex shares with the edge of interest a single node. 
The remaining one depends instead on the state of a 3-node motif, and corresponds to the case where the focal edge belongs to the $2$-simplex. 
In the first two cases, infections occur when the shared node is susceptible and the other nodes of the $2$-simplex are infected, while the other node in the pair can be either susceptible or infected. 
Both the densities of edges in states $(S,S)$ and $(S,I)$ are, hence, affected by this infection process, with a rate that is function of $\beta_{\Delta}$ and the density of 4-node motifs in the aforementioned states. To simplify, we only consider 4-node motifs that only contain a unique 2-simplex.
As shown in Fig.~\ref{fig:4motifs_state}, there are three possible motifs that contribute to this infection process. 
They differ for the number of links (zero, one or two) that connect the node external to the $2$-simplex to the infected nodes of the $2$-simplex.
Hereby, we denote as $\dens{IIS^\Delta S}_0$, $\dens{IIS^\Delta S}_1$, and $\dens{IIS^\Delta S}_2$ the densities of the these motifs in state $(I,I,S,S)$, while we denote as $\dens{IIS^\Delta I}_0$, $\dens{IIS^\Delta I}_1$, and $\dens{IIS^\Delta I}_2$ the density of motifs in state $(I,I,S,I)$. 
Using the clustering coefficient $\phi$, we can write the density of quadruplets of nodes in states $(I,I,S,S)$ and $(I,I,S,I)$ as
\begin{equation}\label{eq:4node_motif_relation}
    \begin{array}{lll}
    \av{IIS^\Delta S} &=& (1-\phi)^2\av{IIS^\Delta S}_0 \\&&+ 2 (1-\phi)\phi \av{IIS^\Delta S}_1 \\&& + \phi^2  \av{IIS^\Delta S}_2 \\ [10pt]
    \av{IIS^\Delta I} &=& (1-\phi)^2\av{IIS^\Delta I}_0 \\&&+  (1-\phi)\phi \av{IIS^\Delta I}_1 \\&& + \phi^2  \av{IIS^\Delta I}_2 \\
    \end{array}
\end{equation}
Next, to evaluate the terms in the model equations that quantify the transitions involving the state of 4-node motifs, we need to calculate the average number of motifs in which the pair of nodes is not part of the $2$-simplex. 
As each node is connected to $k$ links on average, the average number of motifs composed by a $2$-simplex connected to a link is $k(k-1)\phi\delta(k-2)$. 
Indeed, the average number of $2$-simplices connected to a node is $k(k-1)\phi\delta$, while the fourth node of the motif has to be chosen among the $(k-2)$ remaining neighbors of the node. 

We now turn our attention to the three-body infection process involving 3-node motifs. 
In this case, the focal edge is part of the $2$-simplex, and the only possible transition is the one from state $(S,I)$ to state $(I,I)$. 
In fact, both the two other nodes of the $2$-simplex have to be in the infectious state to yield a simplicial contagion, as expressed by the second kinetic equation \eqref{eq:simplicial_SIS_transitions}. 
Similarly to the previous case, the overall contribution present in the model will depend on the infection rate $\beta_{\Delta}$, on the density of $2$-simplices in state $(I,I,S)$, which we denote as $\dens{IIS^\Delta}$, and on the average number of triangles an edge is part of. 
In particular, the latter is given by $2(k-1)\phi\delta=2 k_\Delta/k$, where the factor 2 comes from the fact that each $2$-simplex has two edges pointing to a node. 

Finally, we can write the equations governing the simplicial SIS model at the pair-level
\begin{equation}\label{eq:sis_simplagion_secondorder_normalized}
\begin{array}{ll}
\dot{\langle S\rangle} =& \mu \langle I\rangle - \beta k\langle SI\rangle - \beta_\Delta k_\Delta \langle ISI^\Delta\rangle \\[5pt]
\dot{\langle I\rangle} =& - \mu \langle I\rangle + \beta k\langle SI\rangle  + \beta_\Delta k_\Delta\langle ISI^\Delta \rangle\\[5pt]
\dot{\langle SS\rangle} =& 2\mu\langle SI\rangle - 2\beta(k-1)\langle SSI\rangle \\[5pt]&- 2\beta_\Delta \frac{k_\Delta}{k}(k-2) \langle IIS^\Delta S\rangle\\[5pt]
\dot{\langle SI\rangle} =& \mu \langle II\rangle - \mu \langle SI\rangle + \beta (k-1)\langle SSI\rangle\\[5pt]& - \beta (k-1)\langle ISI\rangle - \beta \langle SI\rangle \\[5pt]&
+ \beta_\Delta \frac{k_\Delta}{k}(k-2)\langle IIS^\Delta S\rangle - 2\beta_\Delta \frac{k_\Delta}{k} \langle ISI^\Delta\rangle\\[5pt]& - \beta_\Delta\frac{k_\Delta}{k}(k-2)\langle IIS^\Delta I\rangle\\[5pt]
\dot{\av{II}} =& - 2 \mu \av{II} + 2 \beta (k-1)\av{ISI} + 2 \beta\av{SI}\\[5pt]& + 4\beta_\Delta \frac{k_\Delta}{k}\av{ISI^\Delta} + 2 \beta_\Delta \frac{k_\Delta}{k}(k-2) \av{IIS^\Delta I}
\end{array}.
\end{equation}

We remark here a property of the model. 
As for the SIS process on networks, in the simplicial SIS contagion recovery is a 1-body mechanism and there are 2-body infections, so the dynamics of $m$-body variables in both the individual-based and the pair-based (exact) models depend on $m$-body and $(m+1)$-body quantities. 
For instance, the dynamics of $\dens{I}$ (1-body variable) depends on $\dens{I}$ itself and on $\dens{SI}$ (2-body). 
In the simplicial SIS process, however, there is a further 3-body infection mechanism at play.
Consequently, the equations governing the evolution of the $m$-body variables in the individual-based and the pair-based models are determined by $(m+2)$-body quantities too. 
Hence, the dynamics of $\dens{I}$ (1-body variable) depends not only on $\dens{I}$ itself and on $\dens{SI}$ (2-body), but also on $\dens{ISI^\Delta}$ (3-body).

Similarly to what we have done for Eqs.~\eqref{eq:pairwise_SIS_MF_exact}, we now close Eqs.~\eqref{eq:sis_simplagion_secondorder_normalized} at the level of pairs. 
We use the approximations for the triplet densities $\av{ABC}$ at the pair level given by Eqs.~\eqref{eq:wedge_closure}, \eqref{eq:kirkwood_closure}, and \eqref{eq:closure_used}. 
For the densities of the $4$-node motifs, we rely on the closures introduced in \cite{house2009motif} (see Eqs.~15), re-adapting them to the simplicial case. 
Specifically, we consider
\begin{equation} \label{eq:quadruplet_closure}
    \begin{array}{l}
         \av{IIS^\Delta S}_0 = \dfrac{\av{SI}^2\av{II}\av{SS}}{\av{S}^2\av{I}^2} \\
        \av{IIS^\Delta S}_1 = \dfrac{\av{SI}^3\av{SS}\av{II}}{\av{S}^3\av{I}^3}\\
        \av{IIS^\Delta S}_2 = \dfrac{\av{SS}\av{SI}^4\av{II}}{\av{S}^4\av{I}^4}\\
        \av{IIS^\Delta I}_0 = \dfrac{\av{SI}^3\av{II}}{\av{S}^2\av{I}^2} \\
        \av{IIS^\Delta I}_1 = \dfrac{\av{SI}^3\av{II}^2}{\av{S}^2\av{I}^4}\\
        \av{IIS^\Delta I}_2 = \dfrac{\av{SI}^3\av{II}^3}{\av{S}^2\av{I}^6}\\
    \end{array},
\end{equation}
which represent all the closures at the level of pair state variables for the 4-node state depicted in Fig.~\ref{fig:4motifs_state}. It is worth noting that this is an approximation, as the 4-node state should depend on the dynamical correlations of the 3-node state variables~\cite{house2009motif}. Moreover, by closing the system at the level of pairs and considering the relation in Eq.~\eqref{eq:kirkwood_closure} to express the state of closed triangles, the conservation relations that are exact for the system before the closure, i.e., up to Eqs.~\eqref{eq:sis_simplagion_secondorder_normalized} ($\dens{S} + \dens{I} = 1$, 
$\dens{SS} + \dens{SI} = \dens{S}$, $\dens{SI}+\dens{II} = \dens{I}$),
are no longer valid~\cite{sharkey2015exact}.

Summing up, Eqs.~\eqref{eq:sis_simplagion_secondorder_normalized} along with the expressions \eqref{eq:wedge_closure}, \eqref{eq:kirkwood_closure}, \eqref{eq:closure_used}, and \eqref{eq:quadruplet_closure} constitute a closed approximation for the pair-based mean-field simplicial SIS model.
Notice that, while the individual-based model in Section \ref{sec:individual-based} can be easily extended to the more general scenario of hypergraphs \cite{de2020social} where correlations among three- and two-body interactions are not considered, the pair-based approximation above only applies to the case of simplicial complexes, which satisfy the inclusion property.


\section{Results}
\label{sec:results}

In this section, we analyze the pair-based approximation of the simplicial SIS model. 
We first present a semi-analytical derivation of the epidemic threshold.  
Then, we compare the predictions of the pair-based and individual-based approximation with the results of stochastic simulations on synthetic simplicial complexes. 
We show that the pair-based simplicial SIS model better predicts not only the epidemic threshold and the nature of the phase transitions, i.e., continuous or discontinuous, but also the temporal evolution of the spreading process.

\subsection{Epidemic threshold for the 
pair-based mean-field simplicial SIS model}

To derive the epidemic threshold, we consider the initial phase of the spreading. At this stage, the population is almost entirely made up of susceptible individuals, and we can determine the condition under which an outbreak can occur by looking at when the density of infected individuals grows. 
By replacing the density of triangles 
$\dens{ISI^\Delta}$ in the second of Eqs.~\eqref{eq:sis_simplagion_secondorder_normalized} by the approximation in 
Eq.~\eqref{eq:kirkwood_closure}, we can write the dynamics of $\langle I \rangle$ as:

\begin{equation}
\begin{array}{ll}
\dot{\langle I \rangle} &= - \mu \langle I \rangle + \beta k \langle SI \rangle + \beta_\Delta k_\Delta \dfrac{\langle SI \rangle^2 \langle II \rangle}{\langle S \rangle \langle I \rangle^2} \\[10pt]
&= \mu \langle I \rangle \left(  \dfrac{\beta k}{\mu} \dfrac{\langle SI \rangle}{\langle I \rangle} +  \dfrac{\beta_\Delta k_\Delta}{\mu} \dfrac{1}{\langle S \rangle} \left(\dfrac{\langle SI \rangle}{\langle I \rangle}\right)^2 \dfrac{\langle II \rangle}{\langle I \rangle} - 1  \right).
\end{array}
\end{equation}
Moreover, by defining the quantity $\mathcal{R}$ as
\begin{equation}
    \begin{array}{l}\label{eq:R}
      \mathcal{R} = \dfrac{\beta k}{\mu} \dfrac{\langle SI \rangle}{\langle I \rangle} +  \dfrac{\beta_\Delta k_\Delta}{\mu} \dfrac{1}{\langle S \rangle} \left(\dfrac{\langle SI \rangle}{\langle I \rangle}\right)^2 \dfrac{\langle II \rangle}{\langle I \rangle} - 1,
    \end{array}
\end{equation}
we can rewrite this equation as
\begin{equation}
\label{eq:condition_from_I}
\begin{array}{ll}
\dot{\av{I}}&= \mu \av{I} \mathcal{R}.
\end{array}
\end{equation}
The quantity $\mu \langle I \rangle$ remains non-negative throughout the entire spreading process. 
Therefore, $\mathcal{R} > 0$ represents the condition under which contagion can occur at the early stage of the process. 
Conversely, if $\mathcal{R} < 0$, the contagion dies out. 
We assume that the values of all the parameters in $\mathcal{R}$, namely $\mu, \beta, 
\beta_\Delta$ and $k_\Delta$ are all known. 
Hence, in order to assess whether an outbreak can occur or not, we only require further information about the quantities $\frac{\langle SI \rangle}{\langle I \rangle}$ and $\frac{\langle II \rangle}{\langle I \rangle}$ at the early stage of the contagion process. Therefore, it is convenient to define the new variables \cite{keeling1999effects,barnard2019epidemic} 
\begin{equation}\label{eq:fast_variables_definition}
    \begin{array}{ll}
         \Pi =\dfrac{\av{SI}}{\av{I}}; & \Psi = \dfrac{\av{II}}{\av{I}},
    \end{array}
\end{equation}
which take finite (possibly non-zero) values even in the limit $t \to 0$, when 
$(\av{S}, \av{I}, \av{SS}, \av{SI}, \av{II}) \to 
(1,0,1,0,0)$, 
and are able to capture the early time correlation between 
susceptible and infected nodes. 
Compared to the dynamics of $\av{I}$, $\av{SI}$ and $\av{II}$, the two new quantities   
$\Pi$ and $\Psi$ 
are \textit{fast variables} \cite{barnard2019epidemic}, meaning that they quickly converge to a quasi-equilibrium state, which we denote as $\left(\bar{\Pi},\bar{\Psi}\right)$.

The equations governing the time evolution of these fast variables, and all the details about their derivation can be found in Appendix A. 
This allows us to study the behavior of $\mathcal{R}$ at the early stage of the contagion.
Specifically, by defining the rescaled infection rates $\lambda = k\beta/\mu$ and $\lambda_\Delta = k_\Delta \beta_\Delta / \mu$, we can express Eq.~\eqref{eq:R} at the disease-free state equilibrium as 
\begin{equation}
    \begin{array}{l}\label{eq:R_diseasefree}
  \mathcal{R} = \lambda\bar{\Pi} + \lambda_\Delta \bar{\Pi}^2 \bar{\Psi} - 1
    \end{array}.
\end{equation}
Still, the complexity of the model makes deriving a closed-form solution for the quasi-stationary states $\left(\bar{\Pi},\bar{\Psi}\right)$ in terms of the model parameters not analytically tractable, leading us to derive these values numerically 
(see Appendix A).

Given the condition for the epidemic outbreak, $\mathcal{R} > 0$, we can express the critical value $\lambda^*$ 
at which the disease-free equilibrium becomes unstable as:
\begin{equation}\label{eq:transcritical_condition}
\lambda^* = \dfrac{1 - \lambda_\Delta \bar{\Pi}^2 \bar{\Psi}}{\bar{\Pi}}.
\end{equation}
This expression shows that 
the stability of the disease-free equilibrium is affected by the presence of higher-order interactions.
Specifically, as their strength increases, i.e. for increasing values of $\lambda_\Delta$, the 
critical value $\lambda^*$ gets smaller. 
This indicates that, in presence of higher-order interactions, the system is more sensitive to an epidemic outbreak, highlighting the critical role of these interactions in the contagion. 
As we will show in the next subsection, this result is agreement with the numerical simulations of the simplicial SIS model on synthetic simplicial complexes. 
In contrast, the individual mean-field approximation of Sec.~\ref{sec:individual-based} predicts an epidemic threshold equal to that of the standard SIS model with only pairwise interactions, i.e., $\lambda^* = 1$ \cite{iacopini2019simplicial}, independently from the value of $\lambda_\Delta$. 
%
Notice that the introduction of fast variables allows us to carry out the linear stability analysis of the disease-free steady state via the numerical evaluation of the leading eigenvalue of the Jacobian matrix of the system given in Eqs.~\eqref{eq:sis_simplagion_secondorder_normalized}. In Appendix B, we show that the leading eigenvalue becomes zero exactly at $\lambda=\lambda^*$, where $\lambda^*$ is given by Eq.~\eqref{eq:transcritical_condition}. However, due to the complexity of the system, the linear stability analysis alone would have prevented us from obtaining a closed-form expression for $\lambda^*$.

\subsection{Comparison with stochastic simulations}

We are now ready to compare the predictions of the pair-based mean-field  approximation with the results of stochastic simulations of the simplicial SIS model on synthetic simplicial complexes generated by 
the random simplicial complex (RSC) model, namely a higher-order generalization of the Erd\"os-Rényi model for random networks \cite{iacopini2019simplicial}. 
We focus on the time evolution $\rho(t) = \av{I(t)}$ of the fraction of infected  individuals, and on its steady-state value $\rho^*$. In particular, we study $\rho^*$ as a function of the model parameters, specifically  the rescaled infection rates $\lambda = k\beta/\mu$ and $\lambda_\Delta = k_\Delta \beta_\Delta / \mu$. 

\begin{figure*}[t!]
    \centering
    \includegraphics[width=\linewidth]{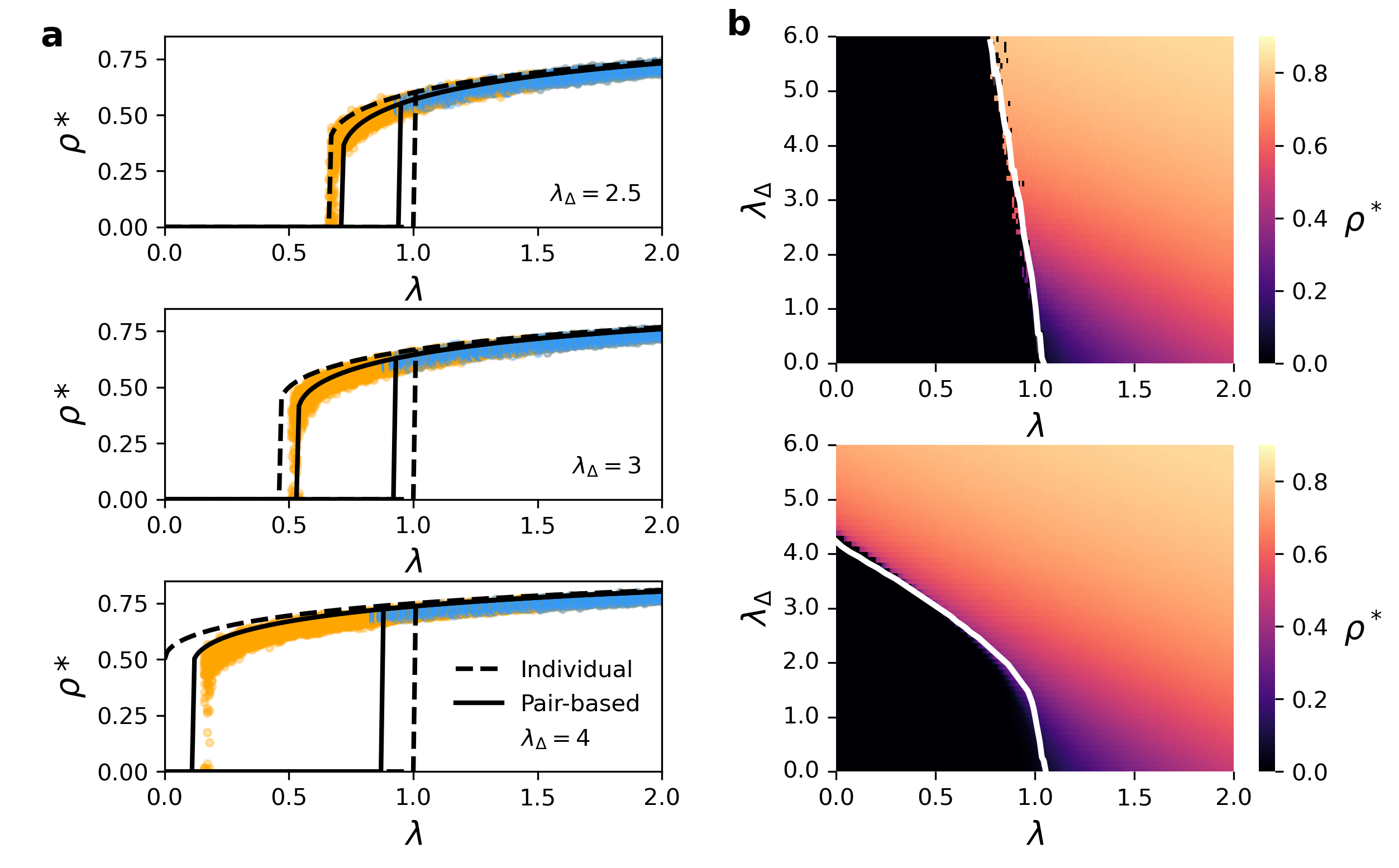}
    \caption{Average fraction $\rho^*$ of infected nodes in the stationary state 
    for the SIS model on a random simplicial complex (RSC) with $N = 2000$ nodes, $k = 20$ and $k_\Delta = 6$. (a) Results of individual (black dashed lines) and pair-based (black solid lines) mean-field approximations are compared to the stochastic simulations for three different values of infectivity 
    $\lambda_\Delta$. The blue dots represent the values of the stationary densities obtained from $M = 100$ iterations of the stochastic simulations with $\rho(0) = 0.001$, while the orange dots represent the stationary densities obtained with $\rho(0) = 0.8$. (b) Phase diagrams reporting 
    $\rho^*$ as a function of the two parameters $\lambda$ and $\lambda_\Delta$. In the upper and lower panel we show the results of the stochastic simulations obtained by starting respectively with initial densities $\rho(0) = 0.001$ and 
    $\rho(0) = 0.8$. The two solid white lines in the upper and lower panel represent respectively the epidemic threshold $\lambda^*$ in Eq. \eqref{eq:transcritical_condition}, and the critical value of $\lambda$  obtained by numerical integration.}
\label{fig:comparison_simulations}
\end{figure*}


\begin{figure*}
\centering
\includegraphics[width=\linewidth]{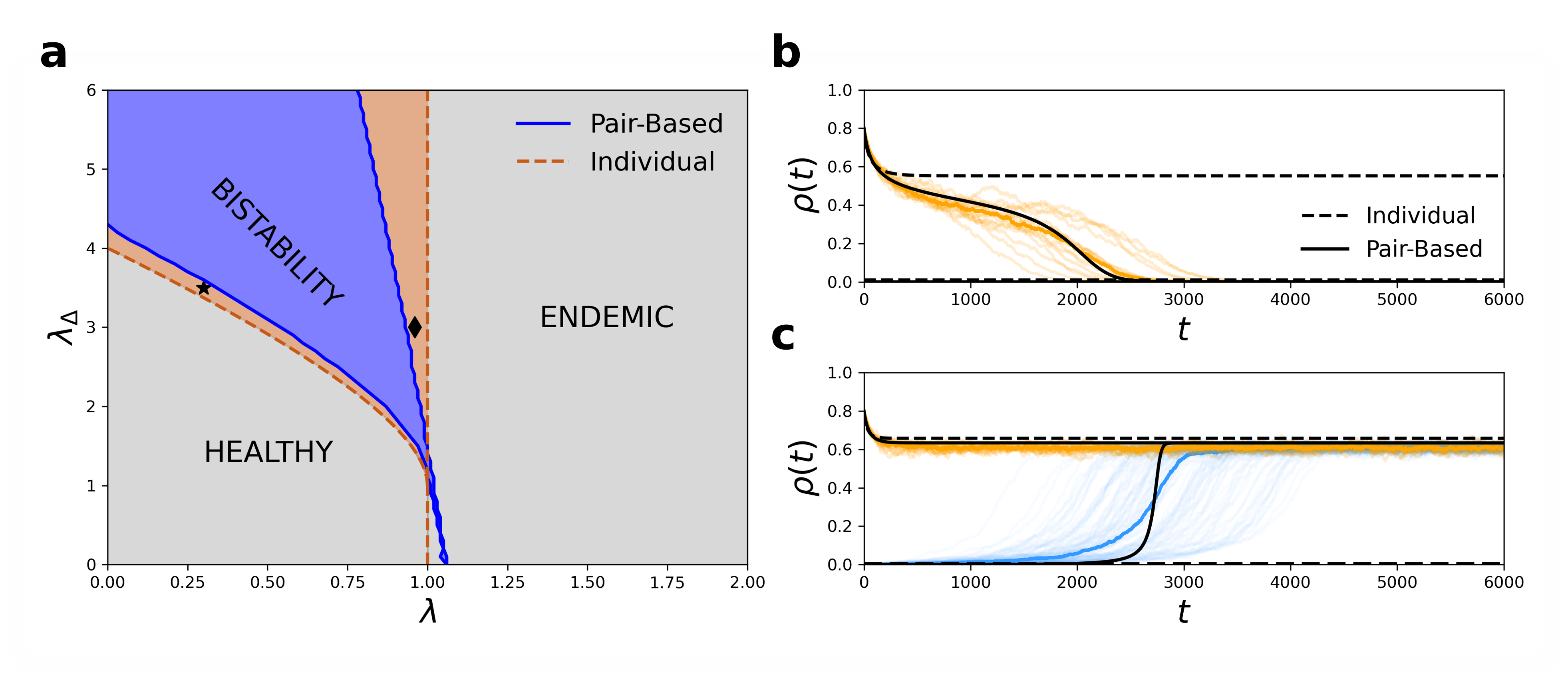}
\caption{Region of bistability of the SIS model and temporal evolution of the fraction of infected individuals.  (a) Phase diagram, in the parameter space ($\lambda$ , $\lambda_\Delta$), highlighting the region of bistability obtained from the individual-based approximation (in orange) and the one derived from the pair-based approximation (in blue). (b) Temporal evolution of the fraction of infected individuals $\rho(t)$ for $(\lambda,\lambda_\Delta) = (0.3,3.5)$, which corresponds to the point marked with a black star in (a), for the individual-based (dashed line) and pair-based (continuous line) approximations. (c) Temporal evolution of the fraction of infected individuals $\rho(t)$ for $(\lambda,\lambda_\Delta) = (0.95,3)$, which corresponds to the point marked with a black diamond in (a), for the individual-based (dashed line) and pair-based (continuous line) approximations. In both panels (b) and (c), for each mean-field model we have illustrated the temporal evolution from two distinct initial conditions, namely $\rho(0) = 0.8$ and $\rho(0) = 0.001$. The colored curves depict $M = 100$ realizations of the spreading process on a RSC, with the colors representing the two different initial conditions of the simulations, namely $\rho(0) = 0.8$ (orange) and $\rho(0) = 0.001$ (blue).} 
\label{fig:fig3}
\end{figure*}

While for the individual-based approximation an analytical evaluation of both the stationary density $\rho^*$ and the condition for the emergence of a saddle-node bifurcation (which marks the onset of the bistability region) can be carried out, for the pair-based approximation this is not possible. Consequently, for the pair-based model, we obtain the stationary density $\rho^*$ through numerical integration of Eqs.~\eqref{eq:sis_simplagion_secondorder_normalized} over a sufficiently long time window, ensuring that the system has reached the steady-state. To investigate the possible onset of a bistable behavior in the pair-based approximation, for each parameter setting, we integrate the model using two distinct initial conditions: a very low fraction of initially infected individuals and a high (close to one) fraction.
Having fixed the initial fraction of infected individuals $\av{I}_0$, from which it follows $\av{S}_0 = 1-\av{I}_0$, we set the initial conditions of the pair variables as
\begin{equation}
    \begin{array}{l}
    \av{SS}_0 = \av{S}_0^2\\
    \av{SI}_0 = \av{S}_0\av{I}_0\\
    \av{II}_0 = \av{I}_0^2
    \end{array}
\end{equation}

Finally, to integrate the pair-based model, we need to evaluate $\phi$, namely the fraction of closed triangles over all triplets in the simplicial complex, which is an input parameter of Eqs.~\eqref{eq:4node_motif_relation}.
For a RSC with $N$ nodes, an average number of $1$-simplices equal to  $k$, and an average number of $2$-simplices equal to $k_\Delta$, the expected number of closed triplets is given by
\begin{equation}
\begin{array}{l}
 \displaystyle \mathbb{E}(\mathcal{T}) \approx  \binom{N}{3}\dfrac{k}{(N-1)} + \binom{N}{3}\dfrac{2k_\Delta}{(N-1)(N-2)}.
\end{array}
\end{equation}
After some algebraic manipulations and using the definition in Eq.~\eqref{eq:clusteringcoeff},
we can approximate the global clustering coefficient $\phi$ for the RSC as
\begin{equation}
    \begin{array}{l}
    \phi \approx \dfrac{k^2 (N-2)}{(N-1)^2 (k-1)} + \dfrac{2 k_\Delta}{k(k-1)}.
    \end{array}
\end{equation}

We carry out stochastic simulations on a RSC with $N = 2000$ nodes, $k \approx 20$ and $k_\Delta \approx 6$, setting $\rho(0)N$ randomly chosen nodes in the infectious state. 
For each setting of the model parameters, we perform $M = 100$ runs, each with a different instance of the RSC model, and evaluate the stationary state $\rho^*$ as the average over the last $100$ values of $\rho(t)$. 
We run the simulations with two different values of $\rho(0)$, namely $\rho(0) = 0.001$ and $\rho(0) = 0.8$.



We now illustrate our results by comparing the stationary density $\rho^*$ obtained with the stochastic simulations, the individual-based, and the pair-based model for different sets of parameters. 
Fig.~\ref{fig:comparison_simulations}(a) shows $\rho^*$ as a function of the rescaled infectivity $\lambda$ for three different values of $\lambda_\Delta$. 
The outcome of the individual-based model is represented by a dashed black curve, while that of the pair-based model by a continuous black line. 
The results of the stochastic simulations with $\rho(0) = 0.001$ are depicted as blue dots, while those with $\rho(0) = 0.8$ as yellow dots.

We observe that both the individual-based and the pair-based approximations predict the existence of an interval of values of $\lambda$ for which the system is bistable, and the transition from the disease-free state, $\rho^*=0$, to an endemic equilibrium, $\rho^* > 0$, is discountinuous. 
However, the pair-based approximation, in general, better predicts the behavior of the stochastic simulations compared to the individual-based approximation. In particular, the individual-based model tends to overestimate the stationary fraction of infected individuals, while the pair-based model is in better agreement with the numerical simulations. Our analysis also shows that the pair-based approximation better identifies the two epidemic thresholds delimiting the region of bistability, while the individual-based model overestimates the width of this region. 
The individual-based model, in fact, underestimates the first threshold, representing critical point for the simulations starting from a high fraction of infected individuals, and, at the same time, overestimates the second threshold, representing the critical point for the simulations starting from a low fraction of infected individuals. 
In particular, the individual-based approximation predicts this second transition at $\lambda = 1$ for any value of $\lambda_\Delta$, whereas the stochastic simulations display a discontinuity at values of $\lambda < 1$ that depend on $\lambda_\Delta$. This behavior is instead correctly reproduced by the pair-based mean-field approximation.
Moreover, the semi-analytical expression of the epidemic threshold in  Eq.~\eqref{eq:transcritical_condition}  accurately predicts the value of $\lambda$ at which the phase transition in the numerical simulations occurs.

We now investigate the behavior of $\rho^*$ as a function of both $\lambda$ and $\lambda_\Delta$, comparing the results of the stochastic simulations on the RSC with the predictions of the pair-based mean field. Fig.~\ref{fig:comparison_simulations}(b) shows the results for $\rho(0) = 0.001$ (upper panel) and $\rho(0) = 0.8$ (lower panel).
The color represents the values of $\rho^*$ as obtained through the stochastic simulations, while the solid white lines denote the transition to a state with $\rho^* \neq 0 $, as predicted by the pair-based approximation. 
Specifically, in the lower panel, representing the case where the initial density of infected nodes $\rho(0)$ is large, the white solid line corresponds to the leftmost threshold of the region of bistability, which we here indicate as $\lambda_c$. 
Instead, in the upper panel, obtained from simulations with a small $\rho(0)$, the white solid line corresponds to the epidemic threshold $\lambda^*$ in Eq.~\eqref{eq:transcritical_condition}. 
We observe that both $\lambda_c$ and $\lambda^*$ decrease as a function of $\lambda_\Delta$, meaning that the three-body interactions boost the spreading process by decreasing the thresholds for the onset of the endemic state. 
In addition, the different ways in which $\lambda_c$ and 
$\lambda^*$ decrease with $\lambda_\Delta$
suggest that the width of the bistable region increases with the intensity of the three-body interactions. 
This dependence of the thresholds on $\lambda_\Delta$ is well reproduced by the pair-based approximation of the simplicial SIS model.




Next, we compare the region of bistability predicted by the individual-based and pair-based approximations.  
Fig.~\ref{fig:fig3}(a) shows the model behavior as a function of $\lambda$ and $\lambda_\Delta$, with the orange and blue areas indicating the region of bistability as predicted by the individual-based and the pair-based approximation, respectively. 
The darker lines represents the thresholds associated to the two approximations.
Finally, the gray area represents the region that in both models corresponds to a single equilibrium, either disease-free $\rho^* = 0$ or endemic $\rho^* > 0$. 
Since the pair-based approximation well predicts the dependence of the critical points on $\lambda_\Delta$, as shown in Fig.~\ref{fig:comparison_simulations}(b), then, Fig.~\ref{fig:fig3}(a) provides another indication that the individual-based model overestimates the width of the region of bistability, by simultaneously underestimating $\lambda_c$ and overestimating $\lambda^*$. 
Overall, these results demonstrate that the pair-based approximation of the simplicial SIS process provides a better description of the spreading process on simplicial complexes compared to the individual-based mean-field approximation.

We conclude this section by illustrating in Fig.~\ref{fig:fig3}(b)-(c) the time evolution of $\rho(t)$ obtained from stochastic simulations on the RSC and from the individual-based and the pair-based approximations for two settings of $(\lambda, \lambda_\Delta)$, namely $(\lambda, \lambda_\Delta) = (0.3,3.5)$ and $(\lambda, \lambda_\Delta) = (0.95,3)$. 
These values correspond to the points of the phase diagram in Fig.~\ref{fig:fig3}(a) marked with a star and a diamond, respectively. 
In both cases, the individual-based approximation predicts the existence of two stable equilibriums, while the pair-based approximation a single one. 
Specifically, in the pair-based model, the steady-state equilibria correspond to a disease-free state for $(\lambda, \lambda_\Delta) = (0.3,3.5)$ and to an endemic state for $(\lambda, \lambda_\Delta) = (0.95,3)$, respectively. 
As shown in Fig.~\ref{fig:fig3}(b)-(c), the time evolution of $\rho(t)$ calculated from the stochastic simulations converges to a single equilibrium point, independently from the system initial conditions, as correctly predicted by the pair-based approximation. 

\section{Discussion and conclusions} 
\label{sec:conclusions}

In this work, we have derived the a pair-based mean-field approximation of the simplicial contagion model and we have compared this approximation to the individual-based approximation.
The key point of our derivation is the introduction of suitable approximations for densities of motifs of three and four nodes, to obtain closed equations at the level of pair of nodes. Consequently, our pair-based approximation can account for the effect of clustering and dynamical correlations that arise during a contagion process in the presence of both two-body and three-body interactions.

For the simplicial SIS model, an evident hierarchy emerges: the state of nodes depends on that of pairs and triples of nodes, the state of pairs depends on triples and quadruples, and so on. 
This dependency on larger motifs is curtailed using closures. 
In our model, variables at the level of triples or quadruples of nodes are approximated by node and pair-level variables. 
This leads to a ``new" system that ideally preserves some of the desirable properties of the original system, particularly the conservation relationships at the level of nodes, namely, pairs, and so on.
Indeed, the conservation relations of the state variables $\langle S \rangle + \langle I \rangle = 1$ and $\langle SS\rangle + 2\langle SI \rangle + \langle II \rangle = 1$ allow to reduce system in Eq.~\eqref{eq:sis_simplagion_secondorder_normalized} to a restricted system of two equations. 
While it has been proved that the pair-based model is exact before closing the system \cite{sharkey2015exact}, Eq.~\eqref{eq:kirkwood_closure} does not preserve triple-level relationships, as it has been shown for pair-based models of SIR processes \cite{taylor2012markovian}.
While an improved closure that preserves the conservation relations could be considered \cite{barnard2019epidemic,house2010impact}, using it for the pair-based approximation of the simplicial contagion model would have increased the system complexity without providing further understanding of the system behavior. 
Moreover, the revised closure does not generally improve the agreement with the numerical simulation of the network process compared to classical closures \cite{kiss2017mathematics}. 
Therefore, to avoid negative state variables and to limit the propagation of this error while integrating the model, we considered the full system of equations instead of the restricted one that uses conservation relations.

In this work, we have also derived an analytical expression for the epidemic threshold in the pair-based mean-field approximation. This allowed us to show that, while in a hypergraph with low correlations between two-body and three-body interactions the epidemic threshold is not affected by the higher-order interactions \cite{burgio2021network}, in a simplicial complex -- a specific type of hypergraph where an interaction among a group of $d$ units necessarily includes all interactions among possible subgroups within that ensemble --
it depends on the strength of higher-order interactions.

Finally, we have compared the predictions of pair-based and individual-based approximations to the numerical simulations of the simplicial SIS model on synthetic random simplicial complexes. 
Our results show that a pair-based approach offers a more accurate description of the higher-order SIS contagion process than an individual-based one.  
This is evident in several different aspects, including the width of the bistability region, the nature of the transition from a disease-free to an endemic state, and the average time evolution of the fraction of infected individuals. 
In particular, the pair-based approximation correctly predicts that the epidemic thresholds depend on the higher-order infectivity.
We notice that, although based on a different approach to derive the mean-field approximation, our conclusion on the dependence of the epidemic threshold from the higher-order infectivity is in agreement with \cite{burgio2023triadic,burgio2021network} that show similar findings for epidemic processes occurring on higher-order structures that adhere to the inclusion property of simplicial complexes. 

The pair-based approximation presented in our work is tailored for systems with higher-order interactions that can be represented as simplicial complexes.  
Hence, our approach is relevant in contexts such as human face-to-face contacts, which are well described in terms of  simplicial complexes \cite{landry2024simpliciality}. 
One direction for future work is to extend our study to the more general case of hypergraphs. 
%
%
%
Another possible direction to explore is the inclusion of both dynamical correlations and degree heterogeneity, so that to extend mean-field models to a larger class of real systems. 

\section*{APPENDIX A: Derivation of fast variable dynamics} \label{appendix:fast_variables}

In this appendix, we provide the details for the calculation of the steady states of the fast variables $\Pi$ and 
$\Psi$, which is the main ingredient in the  derivation of the semi-analytical expression for the epidemic threshold given in Eq.~\eqref{eq:R_diseasefree}.

In the early stages of the spreading process, the fraction of infected individuals, $\av{I}$, and the states of pairs involving infected individuals, namely $\av{SI}$ and $\av{II}$, tend to zero. Consequently, the quantity $\mathcal{R}$ in Eq.~\eqref{eq:R}, which determines whether the contagion occurs or dies out, is ill-defined.
To resolve this issue, we consider the behavior of the system during the early stages of infection, when the population consists almost entirely of susceptible individuals. Under these conditions, the local correlations of infected states surrounding newly infected individuals develop at a much faster rate than the overall densities of susceptible and infected nodes \cite{keeling1999effects}.
Additionally, after an initial transient local correlations are expected to remain relatively constant as long as the clusters of infected nodes developing around the initially infected individuals (invading clusters) remain disconnected from each other. Therefore, we define the variables

\begin{equation}
\label{eq:fast_variables_definition_SM}
\begin{array}{ll}
\Pi = \dfrac{\av{SI}}{\av{I}};&\Psi = \dfrac{\av{II}}{\av{I}},
\end{array}
\end{equation}

The quasi-stationary values of $\Pi$ and 
$\Psi$ can then be used to calculate the condition for the epidemic outbreak, namely the critical value in  Eq.~\eqref{eq:R_diseasefree}.
This approach has been used to analytically determine the epidemic threshold in clustered networks \cite{barnard2019epidemic, keeling1999effects, juher2013outbreak, britton2016network}. 
Here, we extend this methodology to account for the presence of higher-order interactions.

The equation governing the dynamics of $\Pi$ and $\Psi$ are derived from the system in Eq.~\eqref{eq:sis_simplagion_secondorder_normalized}.
To simplify the derivation and the tractability of the model, we consider for both 4-motif states $\av{IIS^\Delta S}$ and $\av{IIS^\Delta I}$, the microscopical configuration where no closed triangles are present in the motif, except the 2-simplex, namely
\begin{equation}
\label{eq:4motif_approximation}
    \begin{array}{l}
\av{IIS^\Delta S} \approx (1-\phi)^2 \av{IIS^\Delta S}_0  \\[5pt]
\av{IIS^\Delta I} \approx (1-\phi)^2 \av{IIS^\Delta I}_0. 
    \end{array}
\end{equation}

Using the standard rule of differentiation and omitting the obvious time dependency, we have
\begin{equation}
    \begin{array}{ll}
         \dot{\av{\Pi}} = \dfrac{\av{SI}'}{\av{I}} - \dfrac{\av{I}' \av{SI}}{\av{I}^2}&  \\[10pt]
         \dot{\av{\Psi}} =  \dfrac{\av{II}'}{\av{I}} - \dfrac{\av{I}' \av{II}}{\av{I}^2}& 
    \end{array}.
\end{equation}

\begin{figure}[t!]
    \centering
\includegraphics[width=0.8\linewidth]{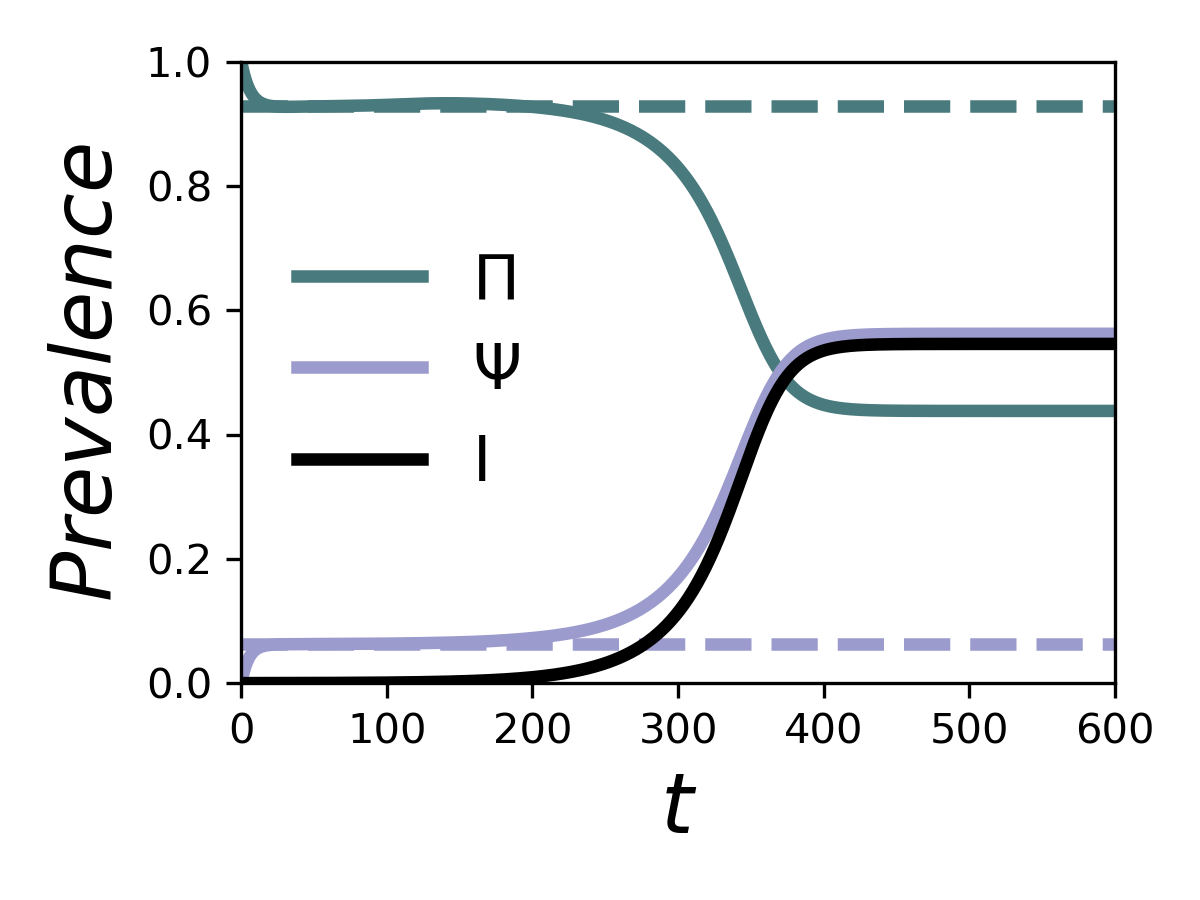}
    \caption{Dynamical evolution of the density of infected individuals  and of the fast variables $\Pi$ (green solid line) and $\Psi$ (violet solid line), as obtained from numerical integration of the system in Eq.\eqref{eq:sis_simplagion_secondorder_normalized}. The dotted lines represent the quasi-stationary values 
    $\bar{\Pi}$ and $ \bar{\Psi}$ as obtained from numerical integration of the system in Eq.\eqref{eq:fast_variables_evolution}.}
    \label{fig:fast_variables}
\end{figure}

Around the disease-free state equilibrium $(\av{S}, \av{I}, \av{SS}, \av{SI}, \av{II}) = (1,0,1,0,0)$, we obtain the following system of equations
\begin{widetext}
\begin{equation}\label{eq:fast_variables_evolution}
    \begin{array}{ll} 
             \dot{\Pi} =& \left[\left(k-1\right)\left(1-\phi\right)- 1\right]\beta \Pi + \mu \Psi  +  \left[ \left(k-1\right)\phi-k\right] \beta \Pi^2 \\[5pt]
            & +\left[-\beta\left(k-1\right)\phi  + \beta_\Delta \frac{k_\Delta}{k} \left(k-2\right)\left(1-\phi\right)^2 - 2\beta_\Delta \frac{k_\Delta}{k}\right] \Pi^2 \Psi - \beta_\Delta k_\Delta \Pi^3 \Psi \\[5pt]
         \dot{\Psi} =& + 2\beta \Pi-  \mu \Psi - \beta k \Pi \Psi+\left[ 2\beta\left(k-1\right) \phi  + 4\beta_\Delta \frac{k_\Delta}{k}\right] \Pi^2 \Psi  - \beta_\Delta k_\Delta \Pi^2 \Psi^2.  
    \end{array}
\end{equation}
\end{widetext}
Since the system in Eq.~\eqref{eq:fast_variables_evolution} is derived around the steady-state equilibrium, the fixed points $\left(\bar{\Pi}, \bar{\Psi}\right)$ of the fast variables coincide with the quasi-stationary states of the full system. 
However, due to the complexity of the system, a closed-form solution for the equilibria as functions of the model parameters is not analytically feasible. Still, we can evaluate $\bar{\Pi}$ and $\bar{\Psi}$ by numerically integrating the system in Eq.~\eqref{eq:fast_variables_evolution} across the entire set of model parameters and plug in these values in the expression of the critical threshold in  Eq.~\eqref{eq:transcritical_condition}. An example of the dynamical evolution of the fast variables $\Pi$ and $\Psi$, obtained through numerical integration of Eq.\eqref{eq:fast_variables_evolution}, is illustrated in Fig.~\ref{fig:fast_variables}, along with the time evolution of the density of infected individuals.

\section*{APPENDIX B: 
Numerical computation of the epidemic threshold}

In this appendix, we show how the critical value $\lambda^*$, as determined by the condition in Eq.~\eqref{eq:transcritical_condition}, is equivalent to the value obtained from the largest eigenvalue of the Jacobian matrix of the system around the disease-free state.

\begin{figure}[t!]
    \centering
\includegraphics[width=\linewidth]{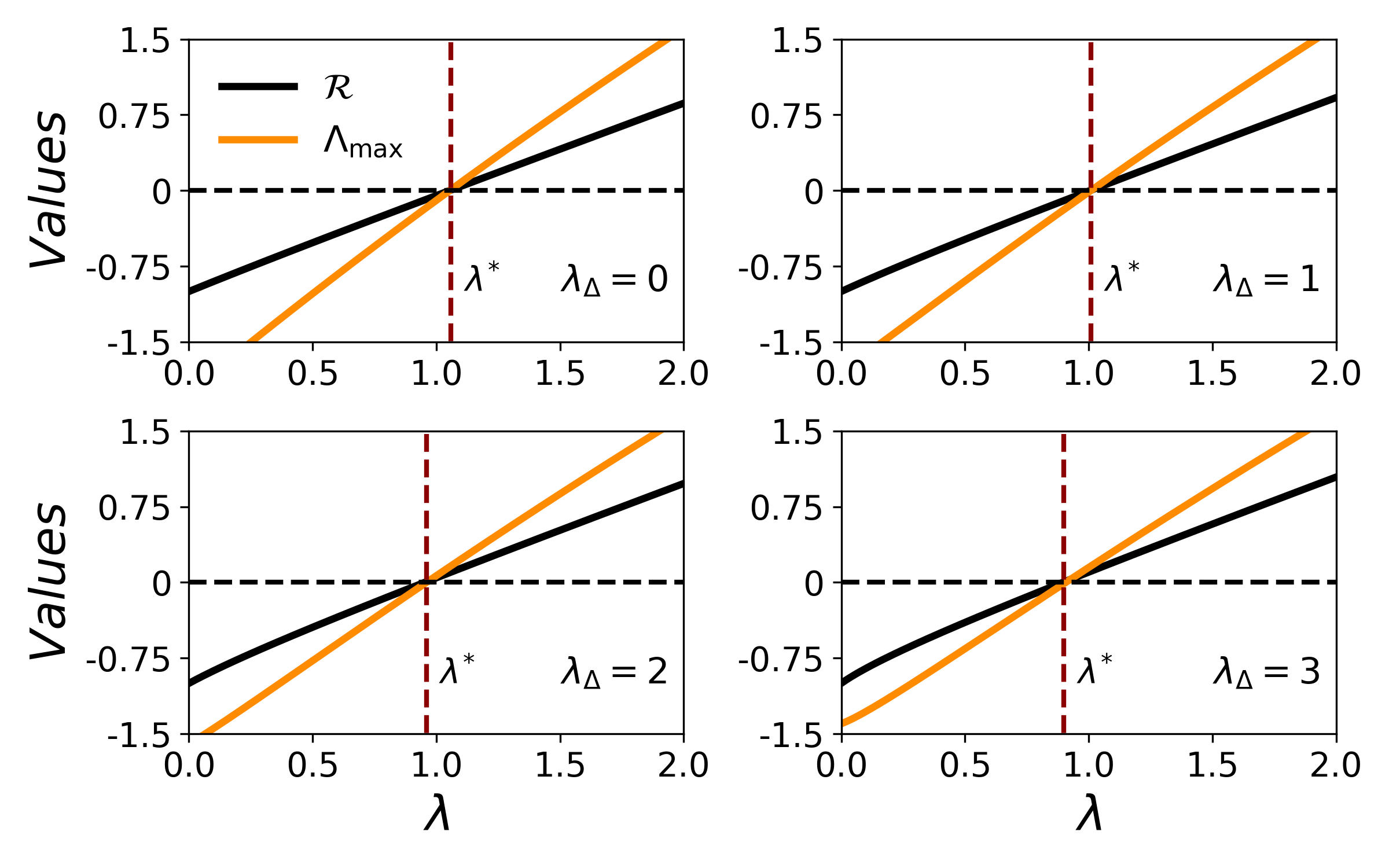}
    \caption{Comparison between the numerical values of $\mathcal{R}$ as given by Eq. \eqref{eq:R_diseasefree} (black solid line) and the maximum eigenvalue of the Jacobian of the restricted system (orange line) as a function of $\lambda$ for four different values of $\lambda_\Delta$. The model parameters are $N=2000$, $k=20$, $k_\Delta=6$, $\mu=0.05$, $\phi \approx 0.05$, and the values of $\bar{\Pi}$ and $\bar{\Psi}$ are obtained through Eq.~\eqref{eq:fast_variables_evolution}. Both conditions are equal to zero at the same critical value, $\lambda^*$.}
    \label{fig:Rvsjacobian}
\end{figure}

To streamline the calculations, we focus on the equations for $\av{S}$, $\av{SI}$, and $\av{SS}$ from Eq.~\eqref{eq:sis_simplagion_secondorder_normalized}. 
We further simplify the derivation employing the approximation in Eq. \eqref{eq:4motif_approximation}.
Finally, we use the closures for the triple states reported in Eq.~\eqref{eq:closed_triples} and for the four-node motif states in Eq.~\eqref{eq:quadruplet_closure}. 
We can thus express the restricted system of equations as

\begin{widetext}
    \begin{equation}
    \begin{array}{ll}
\dot{\langle S\rangle} =& \mu \langle I\rangle - \beta k\langle SI\rangle - \beta_\Delta k_\Delta \dfrac{\av{SI}^2\av{II}}{\av{S}\av{I}^2} \\[10pt]
\dot{\langle SS\rangle} =& 2\mu\langle SI\rangle - 2\beta(k-1)(1-\phi) \dfrac{\av{SS}\av{SI}}{\av{S}} - 2\beta(k-1)\phi \dfrac{\av{SS}\av{SI}^2}{\av{S}^2\av{I}} - 2\beta_\Delta \frac{k_\Delta}{k}(k-2)(1-\phi)^2 \dfrac{\av{SI}^2 \av{II}\av{SS}}{\av{S}^2\av{I}^2}\\[10pt]
\dot{\langle SI\rangle} =& \mu \langle II\rangle - \mu \langle SI\rangle  + \beta(k-1)(1-\phi) \dfrac{\av{SS}\av{SI}}{\av{S}} + \beta(k-1)\phi \dfrac{\av{SS}\av{SI}^2}{\av{S}^2\av{I}} -   \beta (k-1)(1-\phi)\dfrac{\av{SI}^2}{\av{S}} -  \beta (k-1)\phi\dfrac{\av{SI}^2\av{II}}{\av{S}\av{I}^2}\\[10pt]& - \beta \langle SI\rangle 
 + \beta_\Delta \frac{k_\Delta}{k}(k-2)(1-\phi)^2 \dfrac{\av{SI}^2 \av{II}\av{SS}}{\av{S}^2\av{I}^2} - 2\beta_\Delta \frac{k_\Delta}{k}\dfrac{\av{SI}^2\av{II}}{\av{S}\av{I}^2}
 -\beta_\Delta \frac{k_\Delta}{k}(k-2)(1-\phi)^2 \dfrac{\av{SI}^3\av{II}}{\av{S}^2\av{I}^2}.
 \end{array}
\end{equation}
\end{widetext}
The other variables of the system can be obtained considering the conservation of state variables, namely $\av{I} = 1 - \av{S}$ and $\av{II} = 1 - 2\av{SI} - \av{SS}$.

To evaluate the stability of the disease-free state equilibrium, we calculate the Jacobian of the system, denoting
\begin{equation}
\begin{array}{lll}
    f_1(y) = \dot{\av{S}} ;&f_2(y) = \dot{\av{SI}};&f_3(y) = \dot{\av{SS}},
\end{array}
\end{equation}
where $y = (S,SI,SS)$.

We can express the terms of the Jacobian matrix $\mathbf{J}$ around the disease-free state equilibrium $y = (1,0,1)$ as
\begin{widetext}
    \begin{equation}
    \centering
    \begin{array}{ll}
   \dfrac{\partial f_1(y)}{\partial \av{S}}\at[\big]{y = (1,0,1)} =&  - 2 \bar{\Pi}^{2} {\bar{\Psi}} \beta_\Delta k_\Delta - \mu\\[10pt]
    \dfrac{\partial f_1(y)}{\partial \av{SI}}\at[\big]{y = (1,0,1)} =&  2 \bar{\Pi}^{2} \beta_\Delta k_\Delta - 2 {\bar{\Pi}} {\bar{\Psi}} \beta_\Delta k_\Delta - \beta k\\[10pt]
   \dfrac{\partial f_1(y)}{\partial \av{SS}}\at[\big]{y = (1,0,1)} =&  \bar{\Pi}^{2} \beta_\Delta k_\Delta\\[10pt]
   \dfrac{\partial f_2(y)}{\partial \av{S}}\at[\big]{y = (1,0,1)} =&  - 2 \bar{\Pi}^{2} {\bar{\Psi}} \beta \phi \left(k - 1\right) + 2 {\bar{\Pi}}^{2} {\bar{\Psi}} \beta_\Delta \dfrac{k_\Delta}{k}\left(k - 2\right) \left(1 - \phi\right)^{2}  - 4 \bar{\Pi}^{2} {\bar{\Psi}} \beta_\Delta \dfrac{k_\Delta}{k} + \bar{\Pi}^{2} \beta \phi \left(k - 1\right) \\[10pt]
    \dfrac{\partial f_2(y)}{\partial \av{SI}}\at[\big]{y = (1,0,1)} =&  2 \bar{\Pi}^{2} \beta \phi \left(k - 1\right) - 2 \bar{\Pi}^{2} \beta_\Delta \dfrac{k_\Delta}{k}\left(k - 2\right) \left(1 - \phi\right)^{2} + 4 \bar{\Pi}^{2} \beta_\Delta \dfrac{k_\Delta}{k} - 2 {\bar{\Pi}} {\bar{\Psi}} \beta \phi \left(k - 1\right) \\[5pt]& + 2 {\bar{\Pi}} {\bar{\Psi}} \beta_\Delta \left(k - 2\right) \left(1 - \phi\right)^{2} \dfrac{k_\Delta }{k}  -4 {\bar{\Pi}} {\bar{\Psi}} \beta_\Delta \dfrac{k_\Delta}{k} + 2 {\bar{\Pi}} \beta \phi \left(k - 1\right) + \beta \left(1 - \phi\right) \left(k - 1\right) - \beta - 3 \mu \\[10pt]
    \dfrac{\partial f_2(y)}{\partial \av{SS}}\at[\big]{y = (1,0,1)} =&  \bar{\Pi}^{2} \beta \phi \left(k - 1\right) - {\bar{\Pi}}^{2} \beta_\Delta  \left(1 - \phi\right)^{2} \dfrac{k_\Delta}{k}\left(k - 2\right) + 2 \bar{\Pi}^{2} \beta_\Delta \dfrac{k_\Delta}{k} - \mu \\[10pt]
   \dfrac{\partial f_3(y)}{\partial \av{S}}\at[\big]{y = (1,0,1)} =&  - 4 \bar{\Pi}^{2} {\bar{\Psi}} \beta_\Delta  \dfrac{k_\Delta}{k}\left(k - 2\right)\left(1 - \phi\right)^{2} - 2 \bar{\Pi}^{2} \beta \phi \left(k - 1\right)\\[10pt]
   \dfrac{\partial f_3(y)}{\partial \av{SI}}\at[\big]{y = (1,0,1)} =&  4 \bar{\Pi}^{2} \beta_\Delta  \dfrac{ k_\Delta}{k}\left(k - 2\right)\left(1 - \phi\right)^{2} - 4 {\bar{\Pi}} {\bar{\Psi}} \beta_\Delta \dfrac{ k_\Delta}{k} \left(k - 2\right)\left(1 - \phi\right)^{2} - 4 {\bar{\Pi}} \beta \phi \left(k - 1\right) - 2 \beta  \left(1 - \phi\right) \left(k - 1\right) + 2 \mu \\[10pt]
   \dfrac{\partial f_3(y)}{\partial \av{SS}}\at[\big]{y = (1,0,1)} = & 2 \bar{\Pi}^{2} \beta_\Delta  \dfrac{ k_\Delta}{k}\left(k - 2\right)\left(1 - \phi\right)^{2}
    \end{array},
\end{equation}
\end{widetext}
\noindent where we have used the relation given in Eq.~\eqref{eq:fast_variables_definition_SM} to avoid the presence of ill-defined variables.

We focus on the condition at which the largest eigenvalue $\Lambda_\text{max} = \Lambda_\text{max}(\textbf{J})$ of the Jacobian matrix crosses the imaginary axis, namely $\Lambda_\text{max} > 0$.

In Fig.~\ref{fig:Rvsjacobian}, we compare the numerical evaluation of both $\mathcal{R}$, defined in Eq.~\eqref{eq:R_diseasefree}, and $\Lambda_{\text{max}}$ for $k = 20$, $k_\Delta = 6$, $\mu = 0.05$, $\phi \approx 0.05$ and where the values of $\bar{\Pi}$ and $\bar{\Psi}$ are obtained through Eq.~\eqref{eq:fast_variables_evolution}. Despite exhibiting different behaviors, both conditions converge to zero at the same value of $\lambda$. 
This convergence demonstrates the reliability of the relation obtained in Eq.~\eqref{eq:transcritical_condition} for evaluating the epidemic threshold in the pair-based model.


\end{document}